\documentclass[aps,nofootinbib,twocolumn,superscriptaddress,noshowpacs]{revtex4}

\usepackage[dvips]{graphicx}

\usepackage{amssymb}

\newcommand{\ped}[1]{\ensuremath{_{\rm #1}}}

\begin{document}

\title{Andreev reflection in Au/La$_{2-x}$Sr$_{x}$CuO$_{4}$ point-contact junctions:
\\ separation between pseudogap and phase-coherence gap}

\author{R.\@S. Gonnelli}
\email[Corresponding author. E-mail:]{gonnelli@polito.it}
\affiliation{INFM - Dipartimento di Fisica, Politecnico di Torino,
10129 Torino, Italy}
\author{A. Calzolari}
\affiliation{INFM - Dipartimento di Fisica, Politecnico di Torino,
10129 Torino, Italy}
\author{D. Daghero}
\affiliation{INFM - Dipartimento di Fisica, Politecnico di Torino,
10129 Torino, Italy}
\author{L. Natale}
\affiliation{INFM - Dipartimento di Fisica, Politecnico di Torino,
10129 Torino, Italy}
\author{G.\@A. Ummarino}
\affiliation{INFM - Dipartimento di Fisica, Politecnico di Torino,
10129 Torino, Italy}
\author{V.\@A. Stepanov}
\affiliation{P.N. Lebedev Physical Institute, Russian Academy of
Sciences, 117924 Moscow, Russia}
\author{M. Ferretti}
\affiliation{Dipartimento di Chimica e Chimica Industriale,
Universit$\grave{a}$ di Genova, 16146 Genova, Italy}

\begin{abstract}\vspace{1mm}
We made point-contact measurements with Au tips on
La$_{2-x}$Sr$_{x}$CuO$_{4}$ samples with $0.08\leq x\leq 0.20$ to
investigate the relationship between superconducting gap and
pseudogap. We obtained junctions whose conductance curves
presented typical Andreev reflection features at all temperatures
from 4.2~K up to $T_{\mathrm{c}}^{\mathrm{A}}$ close to the bulk
$T_{\mathrm{c}}$. Their fit with the BTK-Tanaka-Kashiwaya model
gives good results if a ($s+d$)-wave gap symmetry is used. The
doping dependence of the low temperature dominant isotropic gap
component $\Delta_{\mathrm{s}}$ follows very well the
$T_{\mathrm{c}}$ vs. $x$ curve. These results support the
separation between the superconducting (Andreev) gap and the
pseudogap measured by
angle-resolved photoemission spectroscopy (ARPES) and tunneling.
\end{abstract}
\maketitle

The possibility of an experimental investigation of the
relationship between superconducting gap and pseudogap in
high-$T\ped{c}$ cuprates directly arises from the hypothesis,
first suggested by G.~Deutscher \cite{Deutscher}, of a really
different nature of these two energy gaps. Nowadays the idea is
becoming more and more accepted that the pseudogap is a property
of the normal state, maybe a precursor of the opening of the
superconducting gap which is due to the achievement of the phase
coherence in the condensate. Thus, different spectroscopic tools
can be used to detect them. Andreev reflection, being strictly
related to the phase coherence, is a probe of the superconducting
state and therefore can be used to measure the coherence gap. On
the contrary, tunneling and angle-resolved photoemission
spectroscopy (ARPES) are expected to be able to detect the energy
gap in the quasiparticle excitation spectrum, even in the absence
of phase coherence.

In the particular case of La$_{2-x}$Sr$_x$CuO$_4$ (LSCO), few
experiments have been performed to investigate the Andreev gap
\cite{ref6b,ref8}, while some tunneling and ARPES evidences exist
supporting a monotonical increase of the low-temperature gap
amplitude at the decrease of doping \cite{ref4,ref6}.

In this paper we present a thorough study of the doping and
temperature dependence of the superconducting gap in LSCO
extracted from the conductance vs. voltage curves of
point-contact junctions between Au tips and LSCO polycrystalline
samples with six different Sr concentrations ranging from
strongly underdoped ($x=0.08$) to slightly overdoped ($x=0.20$).

Details on the sample preparation are given elsewhere
\cite{nostro}. The good quality of the LSCO samples was evidenced
by XRD powder diffraction \cite{Napoletano} and EDS microprobe
analysis. AC susceptibility and resistivity measurements gave
bulk critical temperatures in good agreement with the standard
curve of $T_{\mathrm{c}}$ vs. $x$ for LSCO \cite{Takagi}. The Au
tip was obtained by electro-chemical etching with
HNO$\ped{3}$+HCl of a 0.2~mm diameter Au wire.

Figure~1 shows some representative low-temperature $I$-$V$
characteristics (solid lines) obtained in samples with $x$=0.08
(a) and $x$=0.12 (b), together with the d$I$/d$V$ vs. $V$ curves
(dashed lines).

Figure~2 reports an example of the low-temperature conductance
curves (open circles) for the six values of $x$ here considered,
normalized so that d$I$/d$V$=1 in the normal state, and vertically
shifted for clarity.

\begin{figure}[!]
\begin{center}
\includegraphics[keepaspectratio,width=0.8\columnwidth]{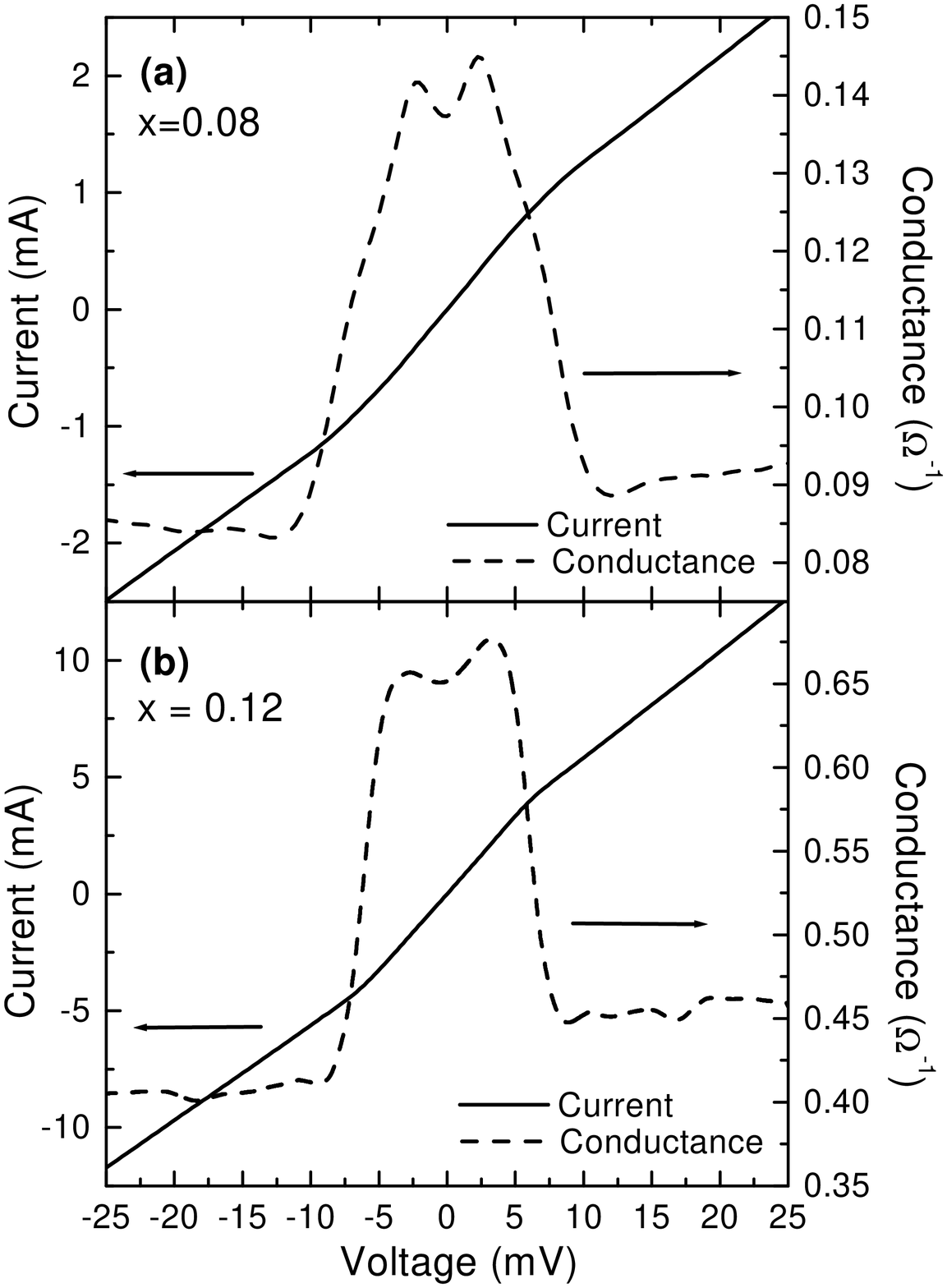}
\caption{ (a) An example of the $I$-$V$ characteristics of a
point-contact junction between the Au tip and a LSCO sample with
$x$=0.08 (solid line), together with the relevant d$I$/d$V$ vs.
$V$ curve (dashed line). (b) The same as (a) but in the case of
$x$=0.12.}
\end{center}
\hspace{5mm}
\includegraphics[keepaspectratio,width=0.75\columnwidth]{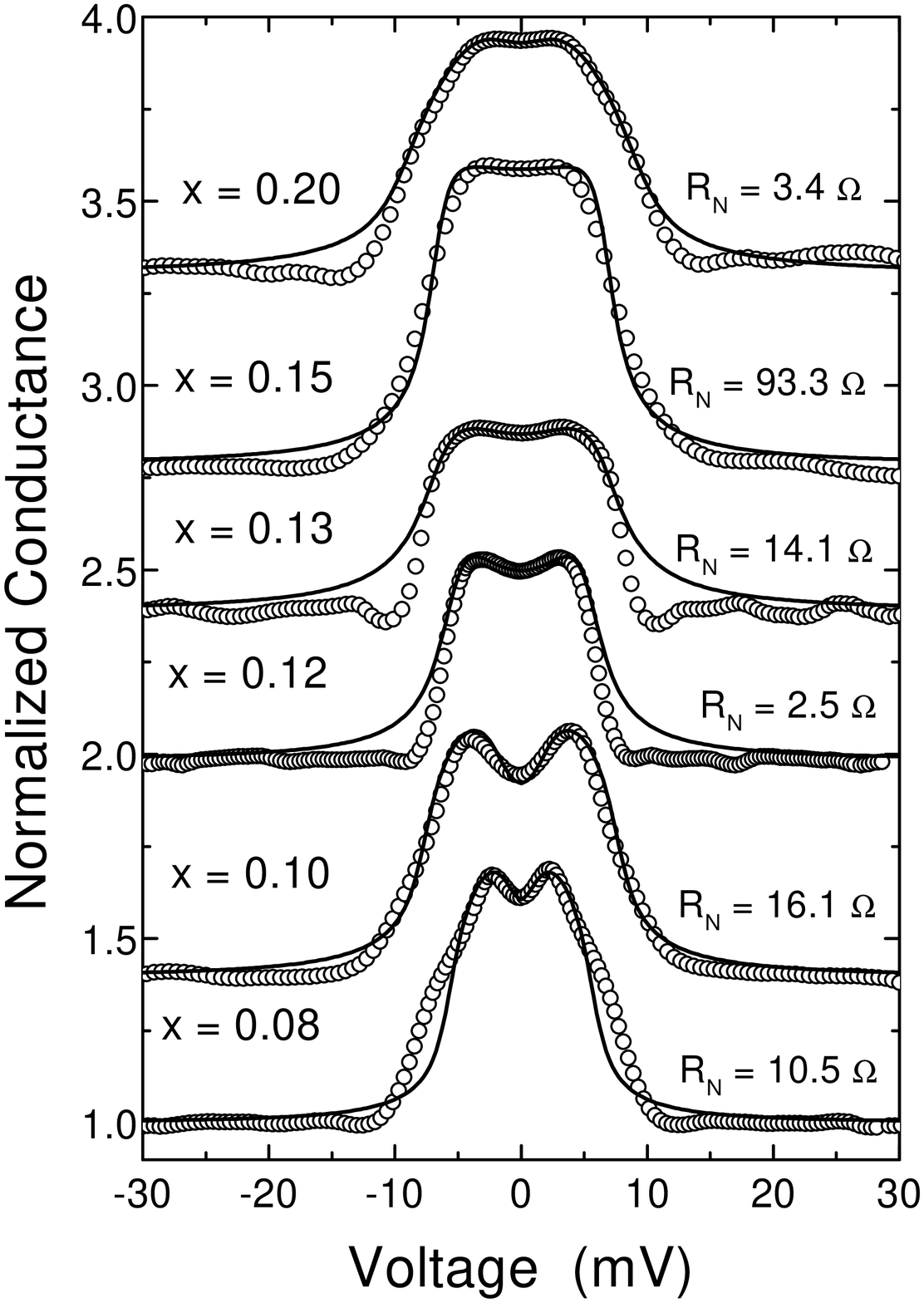}
\vspace{-5mm} \caption{ The normalized conductance curves (open
circles) of Au/LSCO point-contact junctions for various doping
levels ($0.08 \leq x \leq 0.2$) at low temperature (4.22 K $\leq T
\leq$ 5.61 K). The curves are vertically displaced for clarity.
The solid lines represent the best-fit curves calculated by using
the BTK-T-K model. The parameters of each theoretical curve are
indicated in Table~I.}
\end{figure}

These curves and all the others presented in the following were
actually selected among a great number of data sets. We fixed
selection criteria so as to ensure that the measurement was
spectroscopically meaningful and that the result was not affected
by spurious effects. In particular, we required the absence of any
voltage-dependent heating effect and the thermal stability of the
junction. In principle, comparing the conductance curves to those
predicted by the BTK model \cite{BTK} provides by itself a good
probe of the fulfillment of the ideal measurement conditions. Our
experimental curves are indeed fairly similar to the ideal BTK
ones obtained with a finite potential barrier, although their
maximum value is less than that expected and the shape is not
always compatible with a pure \emph{s}-wave gap symmetry. The
oscillations of d$I$/d$V$ at $|V| \gtrsim $ 10 mV are not
``classic'' as well, but are often observed in high-$T\ped{c}$
compounds \cite{ref11b}.

\begin{figure}[t]
\includegraphics[keepaspectratio,width=\columnwidth]{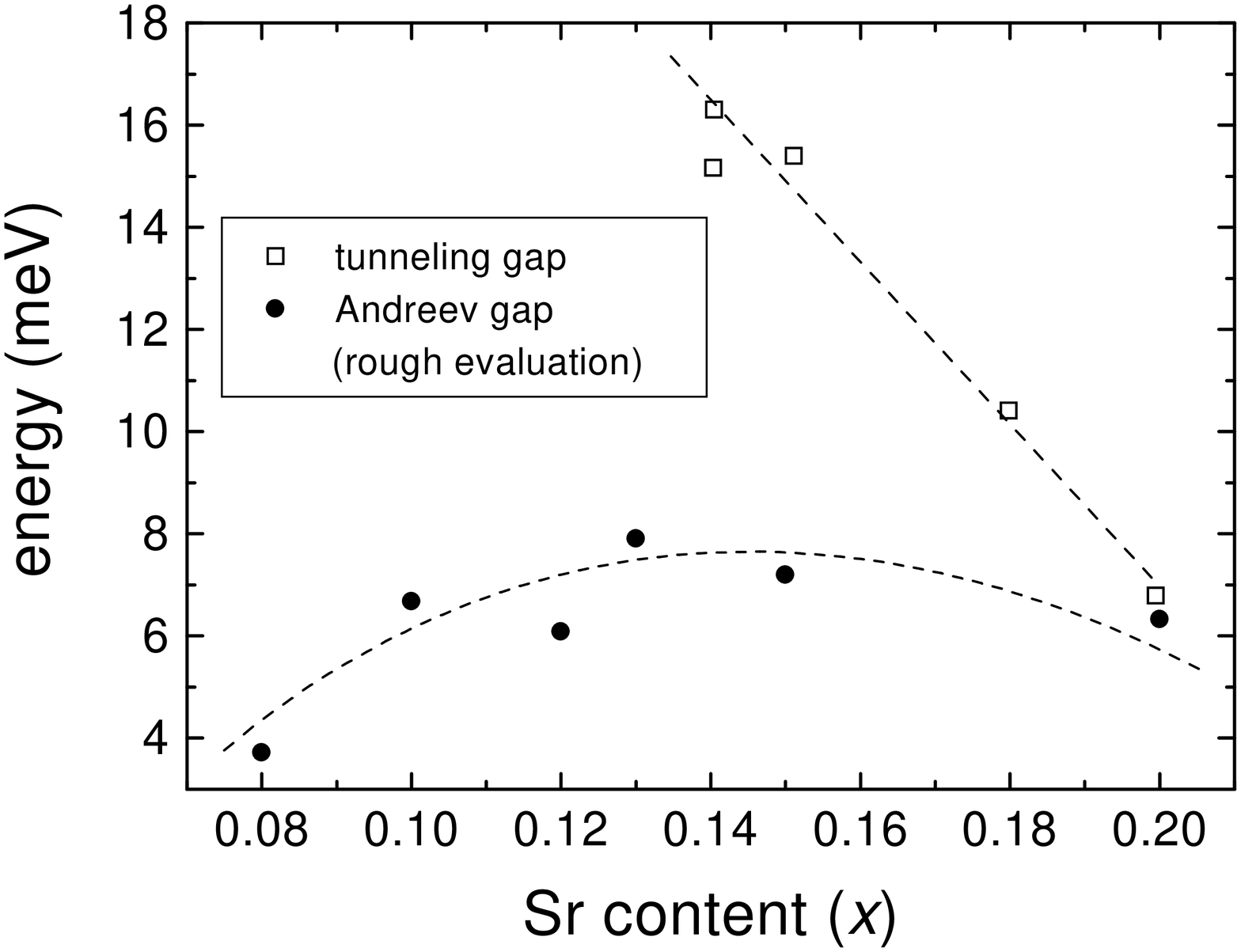}
\caption{Qualitative doping dependence of the superconducting
Andreev gap (solid circles) roughly evaluated as described in the
text. The trend of the superconducting gap is compared to that of
the tunneling gap reported in literature \cite{ref4}.}
\end{figure}

The absence of heating effects is a key point of our discussion.
As well known, point-contact measurements give reliable
spectroscopic information provided that the contact radius $a$ is
smaller than the mean free path $\ell$ in both materials
(ballistic regime). Since any control on the contact dimension is
impossible from the experimental point of view, one usually
evaluates $a$ from the normal-state junction resistance
$R\ped{N}$~\cite{Jansen}. In our case, the values of $R\ped{N}$
reported in Fig.~2 give 146\@\AA$\leq a \leq$\@800\@\AA, whereas
the (evaluated) mean free path $\ell$ ranges from 40 to
70\@\AA~from underdoped to overdoped. These values would rather
indicate that the contact is in the thermal (Maxwell) regime,
characterized by strong heating phenomena. Nevertheless, in the
curves we chose the variation of the normal-state conductance with
bias is very small and well within that expected in the ballistic
regime \cite{Srikanth}. The exceedingly low contact resistances
are thus very likely to be due to the presence of several parallel
ballistic contacts between sample and tip \cite{Aminov}. Anyway,
there's no doubt that the features shown in Fig.s~1 and 2 can only
be produced by Andreev reflection at the S-N interface.

Even at a first glance, the curves in Fig.~2 show that the gap
amplitude increases up to a maximum and then decreases again when
one moves from underdoped to overdoped samples. The simplest way
of evaluating the gap is to identify its edges with the positions
of the conductance maxima. The resulting gap values are shown in
Fig.~3 as a function of doping, together with those measured by
tunneling \cite{ref4}. The two measures almost coincide in the
overdoped region, but differ more and more when the doping is
reduced.

\begin{figure*}[t]
\begin{center}
\includegraphics[keepaspectratio,width=\textwidth]{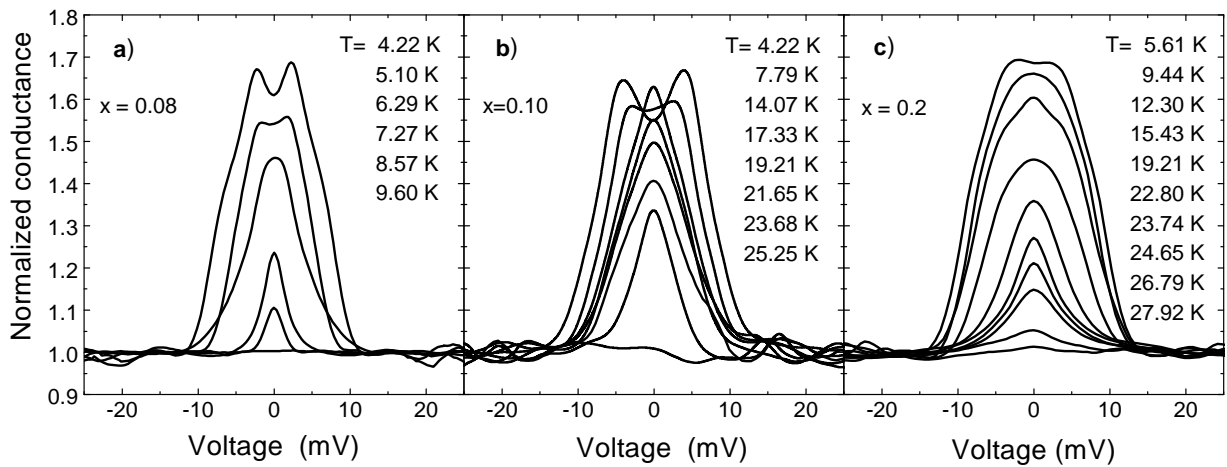}
\vspace{-5cm}\caption{Temperature dependence of the normalized
Andreev conductance in LSCO samples with $x=$ 0.08 (a), 0.10 (b)
and 0.20 (c).}
\end{center}
\end{figure*}

To investigate in greater detail the doping dependence of the
Andreev gap, we compared the experimental curves to the
theoretical ones predicted by the BTK-Tanaka-Kashiwaya model
\cite{BTK-T-K}, in which we also introduced the quasiparticle
lifetime broadening $\Gamma$. If one restricts the analysis to the
low-temperature data of Fig.~2, different possible symmetries of
the order parameter ($s$, $s+d$, $s+$i$d$, \emph{anisotropic~s})
give curves which agree almost equally well with the experimental
ones. Actually, no experimental probes sensible to the gap
symmetry support an \emph{anisotropic~s}-wave symmetry in LSCO,
and therefore we won't consider it.

Further information about which gap symmetry is the most suitable
for the fit can be found in the temperature dependence of the
conductance curves. An example of this dependence for three
different doping contents ($x=0.08$ (a), $x=0.10$ (b) and $x=0.20$
(c)) is shown in Fig.~4, which actually reports for clarity only
some of the curves we measured.

First of all, a result worth to mention is that the Andreev
features always disappear at a temperature $T\ped{c}^{\mathrm{A}}$
close to the bulk $T\ped{c}$ or slightly lower. At $T%
> T\ped{c}^{\mathrm{A}}$, the conductance curves are identical to
those expected in a N-N point-contact ballistic junction
\cite{Srikanth}. The fact that Andreev reflection gives no
evidence of gap above $T\ped{c}$ confirms that it measures the
``true'' superconducting gap, due to the phase coherence.

The information on the gap symmetry we were looking for can be
obtained by fitting the conductance curves in the whole
temperature range (from 4.2~K up to $T\ped{c}^{\mathrm{A}}$) with
the BTK-T-K model. The free parameters, in the general case of
mixed symmetry, are: the isotropic and anisotropic gap components
($\Delta_{\mathrm{is}}$ and $\Delta_{\mathrm{an}}$), the parameter
$Z$ (proportional to the potential barrier height), the lifetime
broadening $\Gamma$ and the angle $\alpha$ between the $a$ axis
and the normal to the S-N interface \cite{BTK-T-K} ($\alpha$ is
unknown because our samples are polycrystalline). Actually, some
constraints reduce the number of adjustable parameters. First,
since $R\ped{N}$ changes very little with $T$, we assumed $Z$ to
be constant and we extracted it from the fits in the various
symmetries of the lowest-temperature curves. We obtained very low
values of $Z$ ($\leq 0.3$), which make the choice of $\alpha$ have
little influence on the values of $\Delta_{\mathrm{is}}$ and
$\Delta_{\mathrm{an}}$ determined by the fit, independently of the
symmetry used \cite{nostro}. Therefore, we could choose $\alpha=0$
without loss of generality. The remaining parameters
$\Delta_{\mathrm{is}}$, $\Delta_{\mathrm{an}}$ and $\Gamma$ were
varied in order to fit the data, but always keeping $\Gamma$ as
small as possible.

A good agreement between the theoretical curves and the
experimental data \emph{in the whole temperature range} is only
obtained if a ($s+d$)-wave gap symmetry is used~\cite{nostro}.
This provides the missing information about the symmetry to be
used for the fit at low temperature, and allows us to refine the
rough evaluation of the doping dependence of the gap sketched in
Fig.~3. The ($s+d$)-wave theoretical curves which best fit the
low-temperature data in Fig.~2 are shown in the same figure as
solid lines. The results are consistent with those obtained in
LSCO by Deutscher \emph{et al.}~\cite{ref6b}. Although the general
symmetry is $s+d$, for some values of $x$ the weight of the $d$
component is zero - that is, the actual symmetry is pure $s$-wave
(but \emph{only} at low temperature). In all cases the $s$-wave
component is dominant and thus is the more representative one.
Also notice that it is a very robust parameter, since its value
would change very little if other gap symmetries were considered.
Table~I resumes all the values of the parameters related to the
curves of Fig.~2.

\begin{table}[t]
\begin{center}
\begin{tabular}{|c|c|c|c|c|c|c|c|}
\hline
  \raisebox{-1.2ex}[0pt][0pt]{Doping} & $\phantom{T}T\phantom{T}$ & $\Delta_{\mathrm{s}}$ & $\Delta_{\mathrm{d}}$ & %
  $\Gamma$ & \raisebox{-1.2ex}[0pt][0pt]{$\phantom{T}Z\phantom{T}$} & %
  \raisebox{-0.3ex}[0pt][0pt]{$\phantom{T}T_\mathrm{c}^\mathrm{A}\phantom{T}$} &
  \raisebox{-0.5ex}[0pt][0pt]{$2\Delta_{\mathrm{s}}$\raisebox{-1ex}{/}%
  \raisebox{-2ex}{$k_\mathrm{B}T_\mathrm{c}^\mathrm{A}$}}
  \\  & (K) & (meV) & (meV) & (meV) & & (K) &
  \\\hline
  0.08 & 4.22 & 3.4 & 2.5 & 0.19 & 0.20 & 9.6 & 8.2 \\ \hline
  0.10 & 4.22 & 4.8 & 3.1 & 0.27 & 0.23 & 25.3 & 4.4 \\ \hline
  0.12 & 4.22 & 5.6 & 0 & 0.92 & 0.18 & 26.0 & 5.0 \\ \hline
  0.13 & 4.22 & 6.8 & 0 & 1.50 & 0.17 & 29.1 & 5.4 \\ \hline
  0.15 & 4.65 & 6.8 & 0 & 0.44 & 0.08 & 35.3 & 4.5 \\ \hline
  0.20 & 5.61 & 6.0 & 3.5 & 1.00 & 0.13 & 27.9 & 5.0 \\ \hline
\end{tabular}
\end{center}
\caption{Best-fit parameters and temperatures for the curves of
Fig.~2.}
\end{table}

In Fig.~5 the doping dependence of the \emph{low-temperature}
$\Delta_{\mathrm{s}}$ (solid circles) obtained from our fit is
compared to those of the ARPES leading-edge shift (LE) recently
determined in LSCO \cite{ref6} (open circles) and of the gap
determined by tunneling measurements (open squares) \cite{ref4}.
The doping dependence of the isotropic gap component
$\Delta\ped{s}$ determined by the fit of our Andreev data confirms
our previous evaluation. In fact, it follows surprisingly well the
$T\ped{c}$ vs $x$ curve (thick solid line) \cite{Takagi}. On the
contrary, both the ARPES LE and the tunneling gap increase
monotonically at the decrease of $x$ and, in the underdoped
region, reach values very larger than those of the superconducting
gap. As a further support to our results, the Andreev gap almost
coincides with the tunneling gap in overdoped samples.

In conclusion, we found that the gap measured by Andreev
reflection spectroscopy in LSCO closes at $T\ped{c}$, and we
obtained a spectroscopic information supporting a mixed $s+d$-wave
symmetry for the order parameter in LSCO. Finally, we found that
the doping dependence of the isotropic component of the
low-temperature Andreev gap clearly follows the $T\ped{c}$ vs. $x$
curve, in contrast with both the tunneling gap and the ARPES LE.

In our opinion, these results support the separation between
pseudopap and phase-coherence (superconducting) gap first claimed
by Deutscher. Within this picture, the pseudogap is a property of
the non-superconducting state of LSCO, independently of its
origin. Although we make no hypothesis about the mechanisms which
govern the opening of the pseudogap, our results are very well
described by some models appeared in literature
\cite{Perali,Benfatto}.

\begin{figure}[t]
\vspace{-3mm}
\includegraphics[keepaspectratio,width=\columnwidth]{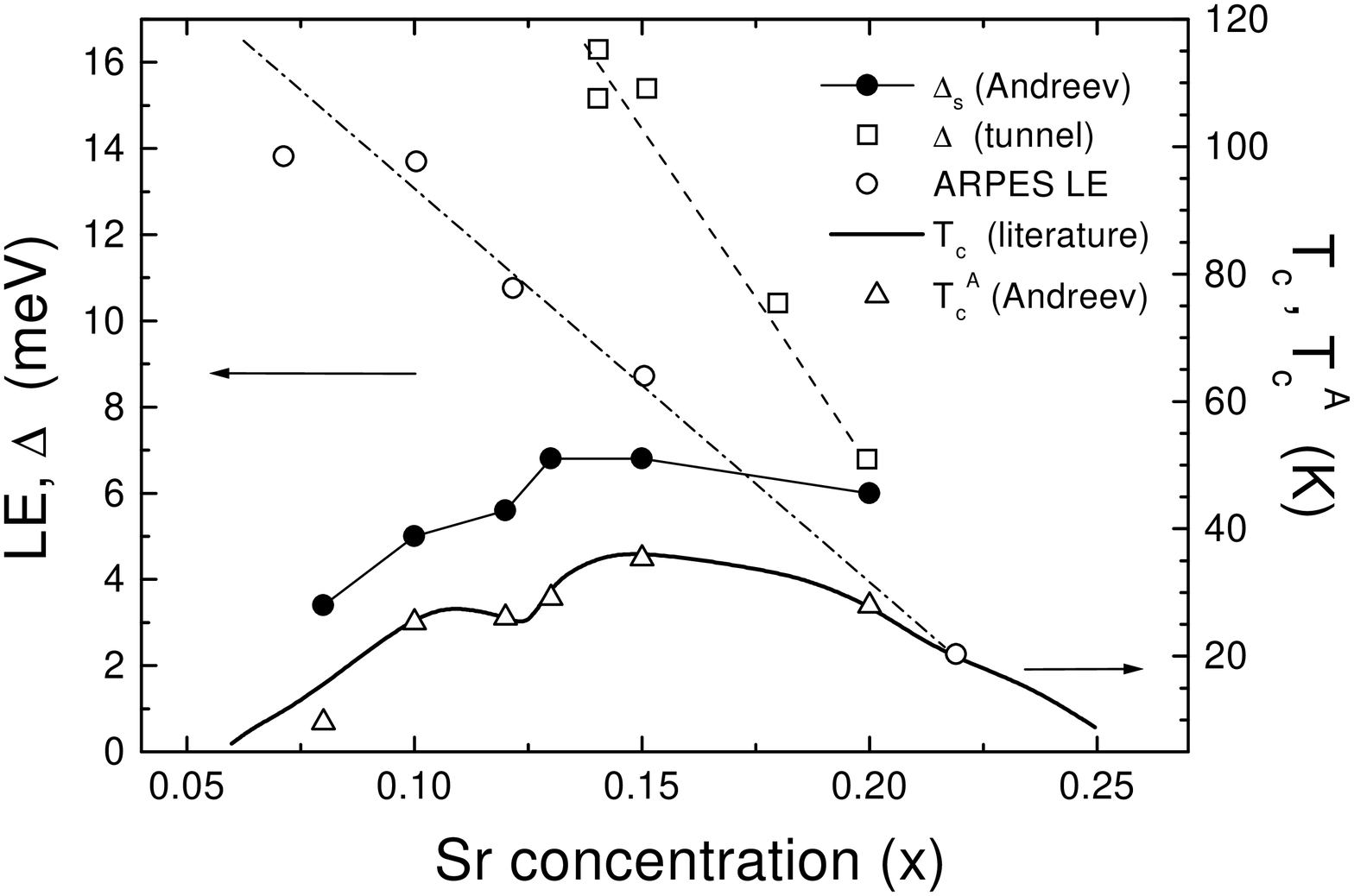}
\caption{Doping dependence of the ARPES leading-edge shift (open
circles, from Ref.\cite{ref6}), of the tunneling gap (open
squares, from Ref.\cite{ref4}) and of our point-contact Andreev
isotropic gap $\Delta_{\mathrm{s}}$ (solid circles) in LSCO. The
temperatures $T_\mathrm{c}^\mathrm{A}$ at which the Andreev
features disappear in our samples are also reported (up triangles)
and compared to the $T_\mathrm{c}$ vs $x$ curve from
Ref.\cite{Takagi} (solid line).}
\end{figure}

This work has been done under the Advanced Research Project
``PRA-SPIS'' of the Istituto Nazionale di Fisica della Materia
(INFM). One of the authors (V.A.S.) acknowledges the partial
support by the Russian Foundation for Basic Research (grant No
99-02-17877) and by the Russian Ministry of Science and Technical
Policy within the program ``Actual Problems of Condensed Matter
Physics'' (grant No  96001).

\end{document}